\documentclass[sigconf, nonacm]{acmart}

\usepackage{hyperref}
\usepackage{xcolor}
\usepackage{amsmath}
\usepackage{amsfonts}
\usepackage{mathtools}
\usepackage{algorithm}
\usepackage{microtype}
\usepackage[noend]{algpseudocode}
\usepackage{colortbl}
\usepackage{booktabs}
\usepackage{tabularray}
\usepackage{makecell}
\usepackage{multicol}
\usepackage{multirow}
\usepackage{graphicx}
\usepackage{subcaption}
\usepackage{cleveref}
\usepackage{tikz}
\usepackage{balance}

\newcommand\vldbavailabilityurl{https://github.com/KaHIP/CompressedStreamingGraphPartitioning}
\newcommand\vldbavailabilityurlsecond{https://github.com/kurpicz/cpi}
\newcommand\vldbpagestyle{plain} 

\def\MdR{\ensuremath{\mathbb{R}}}
\newcommand{\Is}       {\coloneqq}

\newcommand{\gilt}{:}
\newcommand{\setGilt}[2]{\left\{ #1\sodass #2\right\}}
\newcommand{\sodass}{\,:\,}
\newcommand{\framework}{\textsc{StreamCpi}}
\newcommand{\defaultframework}{\textsc{StreamCpi}_{\beta=n,\kappa=1}}

\newcommand{\algDefaultFennel}{\textsc{Fennel-Array}}
\newcommand{\algHashing}{\textsc{Hashing}}
\newcommand{\benchSet}[1]{\multicolumn{9}{c}{#1}}

\newcommand{\instance}[1]{\texttt{#1}}
\newcommand{\instanceSmall}[1]{\small{\instance{#1}}}

\newcommand{\instType}[1]{\small{\textsf{#1}}}
\newcommand{\instMesh}{\instType{Mesh}}
\newcommand{\instCitation}{\instType{Citation}}
\newcommand{\instSocial}{\instType{Social}}
\newcommand{\instWeb}{\instType{Web}}
\newcommand{\instCoPurch}{\instType{CoPur.}}
\newcommand{\instAutSyst}{\instType{AutSys.}}
\newcommand{\instSimilarity}{\instType{Sim.}}
\newcommand{\instRoad}{\instType{Road}}
\newcommand{\instCircuit}{\instType{Circuit}}
\newcommand{\instMatrix}{\instType{Matrix}}
\newcommand{\instGen}{\instType{Gen}}

\begin{document}
\title{Partitioning Trillion Edge Graphs on Edge Devices}

\author{Adil Chhabra}
\affiliation{%
  \institution{University of Heidelberg}
}
\email{adil.chhabra@informatik.uni-heidelberg.de}

\author{Florian Kurpicz}
\affiliation{%
	\institution{Karlsruhe Institute of Technology}
}
\email{kurpicz@kit.edu}

\author{Christian Schulz}
\affiliation{%
	\institution{University of Heidelberg}
}
\email{christian.schulz@informatik.uni-heidelberg.de}

\author{Dominik Schweisgut}
\affiliation{%
	\institution{University of Heidelberg}
}
\email{os226@stud.uni-heidelberg.de}

\author{Daniel Seemaier}
\affiliation{%
	\institution{Karlsruhe Institute of Technology}
}
\email{daniel.seemaier@kit.edu}

\begin{abstract}
Processing large-scale graphs, containing billions of entities, is critical across fields like bioinformatics, high-performance computing, navigation and route planning, among others. 
Efficient graph partitioning, which divides a graph into sub-graphs while minimizing inter-block edges, is essential to graph processing, as it optimizes parallel computing and enhances data locality. Traditional in-memory partitioners, such as \textsc{METIS} and \textsc{KaHIP}, offer high-quality partitions but are often infeasible for enormous graphs due to their substantial memory overhead. Streaming partitioners reduce memory usage to $\mathcal{O}(n)$, where $n$ is the number of nodes of the graph, by loading nodes sequentially and assigning them to blocks on-the-fly. This paper introduces \framework, a novel framework that further reduces the memory overhead of streaming partitioners through run-length compression of block assignments. Notably, \framework\ enables the partitioning of trillion-edge graphs on edge devices. 
Additionally, within this framework, we propose a modification to the \textsc{la\_vector} bit vector for append support, which can be used for online run-length compression in other streaming applications.
Empirical results show that \framework\ reduces memory usage while maintaining or improving partition quality. For instance, using \framework, the \textsc{Fennel} partitioner effectively partitions a graph with 17 billion nodes and 1.03 trillion edges on a Raspberry Pi, achieving significantly better solution quality than \textsc{Hashing}, the only other feasible algorithm on edge devices. \framework\ thus advances graph processing by enabling high-quality partitioning on low-cost machines.
\end{abstract}

\maketitle

\pagestyle{\vldbpagestyle}

\ifdefempty{\vldbavailabilityurl}{}{
\vspace{.3cm}
\begingroup\small\noindent\raggedright\textbf{Source Code Availability:}\\
The open-source software has been made available at \url{\vldbavailabilityurl} and \url{\vldbavailabilityurlsecond}.
\endgroup
}

\section{Introduction}

\label{sec:introduction}

Large-scale graphs, often consisting of billions of entities, are employed across diverse disciplines to model social, biological, navigational, and technical networks. Efficiently processing these graphs can uncover valuable insights -- such as community structures in social networks or pathways in biological systems 
-- determine navigational and recommendation systems and run scientific simulations, among other applications~\cite{alpert1995rdn,alpert1999spectral,catalyuerek1996dis,DellingGPW11,heuvelinecoop,george1973nested,Lau04}. However, processing huge graphs requires extensive computational resources, and concurrently, there is an ever-growing need for scalable graph processing techniques. Several graph-based applications are either preceded by or modeled as \emph{graph partitioning}. In graph partitioning, large graphs are divided into sub-graphs distributed among $k$ blocks of roughly equal size while minimizing the number of inter-block edges. Graph partitioning is crucial for optimizing load balancing in parallel computing, enhancing the performance of search algorithms, and improving data locality in large-scale graph processing. However, graph partitioning is NP-hard~\cite{bourse-2014,Garey1974} and there can be no approximation algorithm with a constant ratio factor for general graphs unless P = NP~\cite{BuiJ92}. Thus, heuristic algorithms \hbox{are used in practice.}

Traditional graph partitioners, like \textsc{METIS}~\cite{karypis1998fast} and \textsc{KaHIP}~\cite{kaffpa}, load the complete graph in memory before partitioning it. These in-memory partitioners produce high solution quality, as they use complete global information to compute partitions, but have a high memory overhead, as they store the entire node set in memory. There is an increasing need to partition huge graphs with less computational resources, especially in scenarios where real-time processing is critical, such as online social network monitoring, fraud detection in financial networks, and dynamic routing in transportation systems. Further, reducing the memory requirements of graph partitioning can lead to significant reductions in monetary costs, as large graphs can be partitioned on small, relatively inexpensive machines. Responding to the need for partitioners with low memory overhead, streaming partitioners load nodes sequentially for immediate assignment to blocks, in lieu of loading the entire node set, thus consuming less memory than \hbox{in-memory partitioners.} 

While streaming partitioners typically yield worse solution quality than in-memory approaches, as they lack complete global knowledge, recently there is a push to enhance their performance by developing more sophisticated streaming techniques that incorporate partial global information~\cite{alistarh2015streaming,awadelkarim2020prioritized,HeiStreamEdge,freight_paper,HeiStream,StreamMultiSection,hoang2019cusp,jafari2021fast,mayer2018adwise,tacsyaran2021streaming,tsourakakis2014fennel}. These partitioners maintain information about blocks assigned to nodes streamed previously to inform future assignments, and subsequently compute a score or objective function for every node for each block $k$, based on information from previous assignments, and then assign that node to the block with the best score. While streaming partitioning using partial global knowledge increases solution quality, it leads to a higher memory overhead, particularly for large graphs, as block assignments are stored for all nodes, typically in a container of size of $\Theta(n)$. 
Thus, there remains \emph{scope to significantly lower the memory overhead} for high-quality streaming graph partitioning, by reducing the memory requirements of storing block assignments. 

\textbf{Contribution.}
In this work, we propose a framework, \framework, that uses run-length compression of block assignments to improve the memory efficiency of streaming graph partitioners. With this framework, we introduce a highly memory-efficient streaming graph partitioner capable of partitioning trillion edge graphs with billions of nodes on edge devices. \framework\ applies problem-tailored run-length data compression techniques, and modified scoring functions to mitigate the memory bottleneck that storing block assignments causes. To enable data compression within this framework, we provide a modification to the state-of-the-art \textsc{la\_vector} learned bit vector to enable append support. This modification supports on-the-fly run-length compression, and is thus suitable for streaming algorithms across various applications.

By reducing memory overhead significantly while maintaining high-quality partitioning, our framework increases the feasibility of graph partitioning, by enabling high-quality partitioning of huge graphs on low-cost machines. 
For example, our partitioner successfully partitions a generated graph with 17 billion nodes and 1.03 trillion edges on a Raspberry Pi with 100\% better solution quality than \textsc{Hashing}, the only other algorithm that can partition this graph on a Raspberry Pi.

\section{Preliminaries}
\label{sec:prelim}

\label{subsec:basic_concepts}

\textbf{Graphs.}
Let $G=(V=\{0,\ldots, n-1\},E)$ be an \emph{undirected graph} with no multiple or self-edges, such that $n = |V|$, $m = |E|$.
Let $c: V \to \MdR_{\geq 0}$ be a node-weight function, and let $\omega: E \to \MdR_{>0}$ be an edge-weight function.
We generalize $c$ and $\omega$ functions to sets, such that $c(V') = \sum_{v\in V'}c(v)$ and $\omega(E') = \sum_{e\in E'}\omega(e)$.
An edge $e = (u, v)$ is said to be \emph{incident} on nodes $u$ and $v$. Let $N(v) = \setGilt{u}{(v,u) \in E}$ denote the neighbors of $v$. 
A graph $S=(V', E')$ is said to be a \emph{subgraph} of $G=(V, E)$ if $V' \subseteq V$ and $E' \subseteq E \cap (V' \times V')$. 
Let $d(v)$ be the degree of node $v$ and $\Delta$ \hbox{be the maximum degree of $G$.}

\textbf{Partitioning.}
Given a number of \emph{blocks} $k \in \mathbb{N}_{\geq 1}$, and an undirected graph with \emph{positive} edge weights, the graph partitioning problem pertains to the partitioning of a graph into $k$ smaller graphs by assigning the nodes of the graph to $k$ mutually exclusive blocks, such that the blocks have roughly the same size and the particular objective function is minimized. More precisely, a \emph{\mbox{$k$-node partition}} of a graph partitions $V$ into $k$ blocks $V_1, \dots, V_k$ such that $V_1 \cup \cdots \cup V_k = V$ and $V_i \cap V_j = \emptyset$ for $i \neq j$. 
The \emph{edge-cut} of a $k$-partition consists of the total weight of the \emph{cut edges}, i.e., edges crossing blocks.
More formally, let the edge-cut be $\sum_{i<j}\omega(E')$, in which $E' \coloneqq $ $\big\{e\in E, \exists (u,v) \subseteq e : u\in V_i,v\in V_j, i \neq j\big\}$ is the~\emph{cut-set}, i.e.,~the set of all \emph{cut edges}.
The \emph{balancing constraint} demands that the sum of node weights in each block does not exceed a threshold associated with some allowed \emph{imbalance}~$\varepsilon$.
More specifically, \hbox{$\forall i \in \{1,\dots,k\} \gilt$ $c(V_i)\leq L_{\max}\Is \big\lceil(1+\varepsilon) \frac{c(V)}{k} \big\rceil$}. 

\textbf{Streaming Partitioning Model.}
Streaming algorithms typically follow a load-compute-store logic. The most commonly used streaming model is the one-pass model where nodes are loaded sequentially and are permanently assigned to blocks. For each node, a score is computed for all blocks of the partition. 
The scoring function indicates how well a node is connected to a block, while maintaining balance constraints, by examining blocks assignments of previously streamed neighbors of the current node.
The node is then permanently assigned to the block with the highest score.

\textbf{Run Length Compression.}
Given a sequence \( S \) of length \( n \), a \emph{run} is a substring of the sequence \( S[i..j] \) where \( 1 \leq i \leq j \leq n \), with \( i \) being the starting index and \( j \) being the ending index of the \emph{run}, satisfying the following conditions:
\begin{enumerate}
	\item \( S[\ell] = S[\ell+1] \) for all \( i \leq \ell < j \).
	\item If \( i > 1 \), then \( S[i-1] \neq S[i] \).
	\item If \( j < n \), then \( S[j] \neq S[j+1] \).
\end{enumerate}
In other words, a \emph{run} is a maximal consecutive sequence of identical characters in \( S \). \footnote{In this paper, our simplified \emph{run} definition, i.e., with fixed period \(p=1\), is sufficient. The \emph{period} is a positive integer \(p\) such that \(S[i]=S[i+p]\), e.g., \texttt{ababab} is a run with period 2.}
The \emph{run length} of a run \( S[i..j] \) is defined as \( j - i + 1 \).
The \emph{run head} of a run \( S[i..j] \) is defined as the character \( S[i] \) (which repeats itself throughout the run).

Then, the \emph{run-length compressed} sequence \(\mathit{RLC}(S)\) is a sequence of tuples consisting of each run's \emph{length} and its corresponding \emph{run head}, for example, \(S=\texttt{aaabaabbba}\) is represented as \(\mathit{RLC}(S)=\langle\texttt{a},3\rangle,\langle\texttt{b},1\rangle,\langle\texttt{a},2\rangle,\langle\texttt{b},3\rangle,\langle\texttt{a}, 1\rangle\).

\subparagraph*{Rank and Select Support.}
The foundation for many compact and compressed data structures is the bit vector, a sequence over the binary alphabet \(\{\mathtt{0}, \mathtt{1}\}\) with rank and select support~\cite{GolynskiRankSelect,GrossiEntropyCompressedTextIndexes,Navarro:compressedindexes}. For a bit vector \(B\) of length \(n\) and \(\alpha \in \{\mathtt{0}, \mathtt{1}\}\), the rank and select operations are defined as follows:
\begin{itemize}
	\item \(\mathit{rank}_\alpha(i) = |\{ j \leq i \colon B[j] = \alpha \}|\) for \(i \in [0, n)\), which counts the number of occurrences of the bit \(\alpha\) in the bit vector up to and including the \(i\)-th position.
	\item \(\mathit{select}_\alpha(x) = \min \{ j \colon \mathit{rank}_\alpha(j) = x \}\), where \(x \leq \mathit{rank}_\alpha(n - 1)\), which finds the position of the \(x\)-th occurrence of the bit \(\alpha\) in the bit vector.
\end{itemize}
These operations can be generalized to alphabets of arbitrary size, extending their utility to various text mining, data compression, and indexing problems.

\section{Related Work}

\label{sec:related_work}
We refer the reader to recent surveys on graph partitioning for relevant literature~\cite{SPPGPOverviewPaper,more_recent_advances_hgp,DBLP:reference/bdt/0003S19}. 
Here, we focus on the research on streaming graph partitioning.

\subparagraph*{Streaming Graph Partitioning.}
Tsourakakis~et~al.~\cite{tsourakakis2014fennel} introduce \textsc{Fennel}, a one-pass partitioning heuristic adapted from the clustering objective \emph{modularity}~\cite{brandes2007modularity}. In this work, we use a modified \textsc{Fennel} scoring function for partitioning on-the-fly. A detailed description of the \textsc{Fennel} objective function is provided in Section~\ref{sec:main_contribution}. \textsc{LDG}, a greedy heuristic proposed by Stanton and Kliot~\cite{stanton2012streaming}, is another streaming partitioning scoring function. Both \textsc{Fennel} and \textsc{LDG} require knowledge of the blocks assigned to the neighbors of the currently streamed node, and the weight of the $k$ partition blocks, i.e., the sum of the weight of nodes assigned to the blocks. As a consequence, these partitioners require $\mathcal{O}(n+k)$ memory to store an array of block assignments of size $n$ and the weights of all blocks. \textsc{ReLDG} and \textsc{ReFennel} are re-streaming versions of \textsc{LDG} and \textsc{Fennel}~\cite{nishimura2013restreaming}, which also require $\mathcal{O}(n+k)$ memory to store block assignments and block weights. In practice, $n >> k$.

An alternative to one-pass streaming is the buffered streaming model, in which a small portion of the graph is stored in-memory. Buffered partitioners like \textsc{HeiStream}~\cite{HeiStream}, which performs multi-level partitioning using a weighted version of the \textsc{Fennel} scoring function, and \textsc{Cuttana}~\cite{cuttana}, which adapts stream ordering to prioritize the partitioning of nodes with the highest number of neighbors previously partitioned, both maintain an array of block assignments for each node. For small enough buffer sizes relative to $n$, the array of block assignments still influences overall \hbox{memory consumption.}

\begin{figure*}
	\centering
	\includegraphics[width=0.9\textwidth]{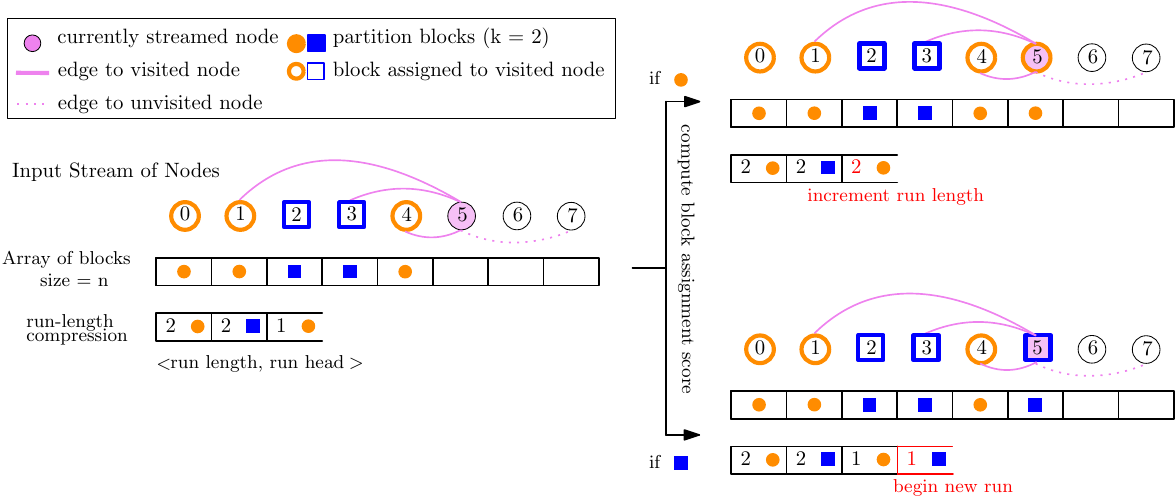} %
	\caption{Schematic of streaming partitioning of a node into orange (circle) and blue (square) blocks. It shows the information held by the partitioner at any instance of streaming: (1) the neighborhood of the current node (node~5), and (2) a container of all nodes' block assignments. Block assignments are used to compute a partitioning score, and node 5 is assigned to the block with the higher score. Run-length encoding of block assignments is shown along with its array representation.}
	\label{fig:streamingrlc} 
\end{figure*}

Unlike the aforementioned streaming graph partitioners, the \textsc{Hashing} heuristic for streaming partitioning does not store block assignments~\cite{stanton2012streaming}. In \textsc{Hashing}, blocks are assigned to nodes by computing a hash function over the node ID: a node $v$ is assigned to block $H(v) = (v \mod k) + 1$. This approach is very fast and requires no memory overhead, but offers extremely poor solution quality, as it produces pseudo-random partitions. Thus, we explore compression of the array of block assignments as a lucrative alternative to reduce the memory consumption of streaming partitioners without sacrificing partition quality.

\subparagraph*{Compact and Compressed Representation of Sequences.}

\label{subsec:rlcrelated}
We refer the reader to implementations of bit vectors with rank and select support, with different space-time trade-offs~\cite{gog2014theory,Kurpicz2022RankSelect,Vigna2008BroadwordRankSelect,ZhouAK2013PopcountRankSelect}. On a bit vector of length $n$, both rank and select queries can be answered in constant time using $o(n)$ \hbox{additional space~\cite{ClarkM1996Select,Jacobson1989LOUDS}}. 

Efficient data compression over bit vectors is generalizable to larger alphabets using wavelet trees~\cite{GrossiEntropyCompressedTextIndexes,NavarroWaveletTreesForAll}, which recursively divide the alphabet $\alpha$, where \(\alpha \in \{\mathtt{0}, \mathtt{1}\}\), by creating a binary tree with nodes corresponding to symbols in the alphabet. Rank queries can be answered in $\mathcal{O}(\log\alpha)$ time by traversing the tree to the leaf level. Using Huffman codes~\cite{Huffman1952HuffmanCodes} to encode symbols in a sequence creates Huffman-shaped wavelet trees~\cite{DBLP:conf/dexaw/BrisaboaCFLP07} that are highly compressible and provide simple random access in time proportional to the length of the code word. In practice, Huffman-shaped wavelet trees can be constructed efficiently~\cite{DinklageEFKL2021PracticalWaveletTrees,DinklageFKT2023SIMDWT} with various techniques to improve query performance~\cite{DBLP:conf/dcc/CereginiKV24,DBLP:conf/dcc/HongBGLZ24}.

While entropy compression techniques like wavelet trees require at least one bit per element of a sequence, dictionary-based compression like (height)-bounded Lempel-Ziv compression algorithms and block trees circumvent this limitation~\cite{BelazzouguiCGGK2021BlockTrees,DBLP:conf/dcc/KreftN10,DBLP:journals/corr/abs-2403-09893}. Block trees require only about 0.1 bits per element on highly repetitive sequences; however, they are orders of magnitude slower to construct than \hbox{wavelet trees~\cite{KopplKM2023LPFBlockTrees}}.

Alternatively, compressed self-indexes, like the \textsc{FM-index}~\cite{FerraginaM2000FMIndex} and the \textsc{r-index}~\cite{GagieNP2020FullyFunctionalRIndex} based on the Burrows-Wheeler transform~\cite{BurrowsW1994BWT} facilitate pattern-matching queries and random access on a compressed sequence, without requiring additional memory. In this work, however, we focus on other representations that support random access on variable bit-length codes, as we do not require more \hbox{sophisticated queries}.

Bit vectors can themselves be compressed, which is space efficient when the sequence encoded in the bit vector is sparse, i.e., it contains few 1s. Theoretically, the current state-of-the-art is by P\v{a}tra\c{s}ku~\cite{Patrascu2008Succincter}. However, it is not practically implementable due to its complexity.  \textsc{RRR}~\cite{RamanRS2007RRR} provides a simple indexable dictionary representation of a sequence that supports rank and select operations using \textit{most-significant-bit first} bucketing. Alternatively, the \textsc{la\_vector} bit vector, as proposed by Boffa et al.~\cite{BoffaFV2021LearnedBitVector}, represents the compressed bit vector as points on a Cartesian plane, and implements rank and select operations through the use of linear approximations to ‘learn’ the distribution of points on the plane. In this work, we use the \textsc{la\_vector} bit vector instead of \textsc{RRR} as its space-time trade-off is more suitable for our application of streaming \hbox{graph partitioning.} 

\section{\textsc{\textmd{StreamCpi}}: A Framework for Memory Optimization in Streaming Partitioning}
\label{sec:main_contribution}

In this section, we introduce our framework \framework\ for reducing the memory consumption of streaming partitioners by \textbf{C}ompressing the array of block assignments (\textbf{P}artition \textbf{I}ndices) used by such partitioners. In particular, we apply run-length data compression to encode runs of repeating block assignments generated by the streaming partitioner on-the-fly. We thus propose a novel \hbox{(semi-)dynamic} compression vector that functions as a drop-in replacement to standard arrays, like std::vector in \textsc{C++}, used to store block assignments in streaming graph partitioners. In this work, we build \framework\ with a modified \textsc{Fennel} partitioning scoring function, the details of which are provided below. 
Principally, our framework can be inserted as a subroutine to reduce the memory footprint in any streaming graph partitioning tool that needs to store block IDs in a \hbox{vector of size $\Theta(n)$}.

\textsc{Fennel}~\cite{tsourakakis2014fennel} assigns a node $v$ to the block $V_i$ that maximizes the \textsc{Fennel} gain function \hbox{$|V_i\cap N(v)|-f(|V_i|)$}, where $f(|V_i|)$ is a penalty function to respect a balancing threshold. The authors define the balancing penalty \hbox{$f(|V_i|) = \alpha \gamma \cdot |V_i|^{\gamma-1}$}, in which~$\gamma$ is a free parameter and $\alpha = m \frac{k^{\gamma-1}}{n^{\gamma}}$. After parameter tuning, the authors set $\gamma=1.5$. The pseudocode of the algorithm is given in Algorithm~\ref{alg:fennel}. 
As described in Section~\ref{sec:related_work}, the memory efficiency of \textsc{Fennel}’s scoring function, like that of other streaming partitioning scoring functions, is limited by the need to store an array of block assignments $A_{part}$. This array, $A_{part}$, has size $\Theta(n)$, and stores the block assigned to each node in its corresponding index.  
In \textsc{Fennel}, computing $|V_i\cap N(v)|$ requires knowing which blocks $A_{part}[w]$ were assigned to the neighbors $w$ of node $v$. Any neighbors that have not been streamed yet are ignored. 
Additionally, computing the penalty \hbox{$f(|V_i|) = \alpha \gamma \cdot |V_i|^{\gamma-1}$} requires knowing the size of each block $V_i$. These are also stored in an array of size $\Theta(k)$. 

The overall memory complexity of \textsc{Fennel} is thus \hbox{$\mathcal{O}(n+k)$}, to store block assignments and block sizes respectively. Typically, the number of nodes $n$ of the graph is much larger than the number of blocks $k$ in the partition, and thus, the array of block assignments $A_{part}$ dominates memory consumption. In \framework, we replace $A_{part}$ with a run-length compressed vector to save space. More specifically, we compress runs of block assignments by storing the \emph{head} of the run and its \emph{length} as defined in Section~\ref{sec:prelim}.
For example, \texttt{1,1,1,1,2,2,3,3,4,5} is stored as \(\langle\mathtt{1},4 \rangle,\langle\mathtt{2}, 2 \rangle,\langle\mathtt{3},2\rangle,\langle\mathtt{4},1\rangle,\langle\mathtt{5},1\rangle\). A schematic of using naive run-length compression with streaming partitioning is shown in Figure~\ref{fig:streamingrlc}. 

To support streaming with access to block assignments, the run-length compressed vector must support append (for storing block assignments of nodes as they are streamed) and query (for score computation). Thus, we propose a (semi-)dynamic compression vector that can perform run-length compression on-the-fly using a bit vector that can perform queries in constant time, and supports append. Next, we provide the details of this data structure in Section \ref{subsec:cpi}. To improve this approach further, we present a batch-wise compression model to speed up append and query times in Section \ref{subsec:batch_cpi}, and offer a simple modification of the \textsc{Fennel} scoring function in Section~\ref{subsec:kappa_modification} for greater memory reduction by encouraging longer runs of block assignments. Finally, in Section~\ref{subsec:expq}, we discuss an alternative approach for the memory optimization of streaming partitioners by using an external-memory priority queue with time-forward processing to store and access block assignments.

\begin{algorithm}[t]
	\begin{algorithmic}[1]
		
		\State \textbf{Input:} Graph $G = (V, E)$, number of blocks $k$
		\State \textbf{Output:} Partition of nodes into $k$ blocks
		
		\State Init $k$ empty blocks $V_1, V_2, \ldots, V_k$ \Comment{\textcolor{gray}{$\Theta(k)$ memory}}
		\State Init array $A_{part}[v] = -1$ $\forall v \in V$ \Comment{\textcolor{gray}{$\Theta(n)$ memory}}%
		
		\For{\textbf{each} $v \in V$}
		\State best block $b^* \leftarrow -1$, best score $S^* \leftarrow -\infty$
		\For{\textbf{each} block $b$} %
		\State $\text{gain} \leftarrow |V_b\cap N(v)|$
		\State $\text{penalty} \leftarrow \alpha \gamma \cdot |V_b|^{\gamma-1}$
		\State $S_b \leftarrow \text{gain} - \text{penalty}$
		\If{$S_b > S^*$}
		\State $b^* \leftarrow b$, $S^* \leftarrow S_b$
		\EndIf
		\EndFor
		\State $A_{part}[v] \leftarrow b^*$, $|V_{b^*}|$+=1
		\Comment{\textcolor{gray}{Assign $v$ to block $V_{b^*}$}}
		\EndFor
	\end{algorithmic}
	\caption{\textsc{Fennel} Partitioning Algorithm}
	\label{alg:fennel}
\end{algorithm}

\subsection{Run-Length Compression of Block Assignments}
\label{subsec:cpi}
We propose a (semi-)dynamic \emph{run-length} compressed vector to store and access block assignments during streaming partitioning, in place of a standard array container. In our implementation, we use a bit vector with rank and select support to perform run-length compression. More specifically, we utilize the \textsc{la\_vector} bit vector~\cite{BoffaFV2021LearnedBitVector}, modified to support the append function, which offers efficient random access and rank queries through learning and adapting to data regularities. We select \textsc{la\_vector} over alternatives because it offers a good trade-off between (rank) query performance and compression ratio on a diverse set of inputs, as generated by the streaming partitioner for different graph types and $k$ values.

\begin{figure}
	\centering
	\includegraphics[width=0.45\textwidth]{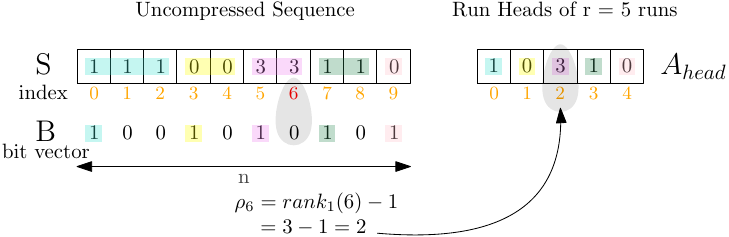} %
	\caption{Run-length compression of a sequence $S$ using a bit vector $B$ and an array of run heads $A_{\text{head}}$. The sequence and bit vector have size $n$. The bit vector contains 1-bit if a new run starts at the corresponding position in $S$, or 0-bit otherwise. In practice, the bit vector is itself compressed. Computing $rank_1(i)$ of an index $i$ returns the number of 1's in the bit vector that occur before index $i$ in $B$. Here, $i=6$ and \hbox{$\rho_6 = rank_1(6) - 1$} returns the position of the run in $A_{\text{head}}$ that $i=6$ belongs to, i.e, \hbox{$A_{\text{head}}[\rho_6]$ = $A_{\text{head}}[2]$ = 3 = $S[6]$}.}   
	\label{fig:bitvec}
\end{figure}

We begin by describing how a bit vector with rank and select support can be used to perform run-length compression of a sequence. Let $r$ be the number of runs contained in the sequence $S$ that we wish to compress. First, we store the run heads of runs in $S$ in an array $A_{\text{head}}$ of length $r$. To match indices of the elements in $S$ to their corresponding run head in $A_{\text{head}}$, we utilize a bit vector $B$ with rank support. The bit vector $B$ has length $n$ and contains a 1-bit at the same index as the starting index of each run in $S$, and a 0-bit otherwise, as shown in Figure~\ref{fig:bitvec}. The bit vector thus contains~$r$~1-bits. 
To determine the run head of the $v$-th element in the uncompressed sequence $S$, we use a single rank query on the bit vector to identify the index of the run head in $A_{head}$:
$\rho_v = \mathit{rank}_1(v) - 1.$
This process is depicted in Figure \ref{fig:bitvec} for $v=6$. Note that each start of a run is marked by a 1-bit in $B$. Therefore, the number of 1-bits up to and including position \(v\) corresponds to the run in which position $v$ resides. Finally, we return the corresponding run-head $A_{\text{head}}[\rho_v]$. 

Storing the bit vector $B$ plainly in memory requires $n$ bits: to save more space, the bit vector is compressed as well. In this work, we use the \textsc{la\_vector} compressed bit vector for more efficient query and space saving. 
In \textsc{la\_vector}, the compressed bit vector is represented by piece-wise linear approximations and some correction terms. 
More specifically, instead of representing the bit vector directly, \textsc{la\_vector} maps each element of the uncompressed sequence to a point in the Cartesian plane. Then, it attempts to build a function that fits these points and can thus be queried to recover the original element. However, finding a function that fits all points can be challenging, hard to compute, or occupy too much space. Thus, points of the plane are split into segments, \(s_1,\dots,s_\ell\), that are each represented by their own linear functions~\(f_j\)~for~\(j\in[1,\ell]\) using Piecewise Linear Approximation. 
For each segment, only the linear function and some correction terms are stored, making \textsc{la\_vector} very space efficient. Further, \textsc{la\_vector} supports rank queries in time proportional to the $\log$ of the number of segments constructed to represent the sequence, which is very fast in practice~\cite{BoffaFV2021LearnedBitVector}. 

However, \textsc{la\_vector} only considers the static compression case, that is, it constructs a compressed bit vector for a sequence that is already built and can not be modified. However, while streaming, we build the sequence of block assignments on-the-fly, and therefore require the ability to append to the bit vector.  
Therefore, as an additional contribution of independent interest, we modify \textsc{la\_vector} to be \hbox{(semi-)dynamic} by adding support \hbox{for an append operation.} 

To incorporate append support in \textsc{la\_vector}, we periodically decompress the last segment of the compressed \textsc{la\_vector} representation built so far, append new elements to it, and then compress it again. However, doing this for every append operation would be very time consuming. Thus, we introduce a buffer of size $\delta$ that stores the starting positions of the most recent $\delta$ uncompressed runs. Once the buffer is full, the last segment of the \textsc{la\_vector} representation is decompressed, and all starting positions are appended to the representation in a batch. This allows our \hbox{(semi-)dynamic} compressed bit vector to have size, i.e., memory consumption, close to the static version, while reducing the time taken to append to it. 

However, our \hbox{(semi-)dynamic} compressed bit vector still has some shortcomings. Firstly, compared to using a standard array, append and query times are still longer. Secondly, in the case of streaming partitioning, when compressing block assignments on-the-fly, the efficiency of compression largely depends on the order in which nodes are streamed, the locality inherent in the graph being partitioned, and consequently, if the block assignments produced by the partitioner exhibit long or short runs. Thus, an efficient run-length compressed data structure can significantly reduce memory consumption \textit{if} the generated partition data contains many repeated block assignments for consecutive nodes.
We propose two modes for \framework\ to address these shortcomings in the upcoming subsections, namely batch-wise compression to improve runtime, and a \textsc{Fennel} modification to encourage block assignments \hbox{with longer runs.}

\subsection{Batch-Wise Compression}
\label{subsec:batch_cpi}
Streaming partitioning scoring functions, like \textsc{Fennel}, require frequent access to block assignments of previously streamed nodes. For every node $v$, we have up to $\lvert N(v) \rvert$ such queries. While standard array containers have random access times of $\mathcal{O}(1)$, our compression vector requires query time proportional to the number of segments in its bit vector representation as detailed in Section~\ref{subsec:cpi}. The number of segments in this representation is likely to be higher for an input of larger size. Thus, to speed up query times from our compression vector, we split the input to be compressed in batches.
In particular, instead of using one compression vector responsible for compressing block assignments for all $n$ nodes, we create $n/\beta$ smaller compression vectors, where $\beta$ is an input parameter specifying the number of nodes whose block assignments we wish to store in a single compression vector. We then insert and \hbox{access block assignments as follows.} 

Let the nodes of the graph be given IDs \hbox{$v \in \{0,1,...,(n-1)\}$}. We append the block assignment of node $v$ to the $j$-th compression vector where $j = v / \beta$ (positive integer division, rounded down). Subsequently, since we stream and append blocks assigned to nodes in monotonically increasing order of node IDs, we can retrieve the block assigned to $v$ from the $v \% \beta$ position of the $j = v / \beta$-th compression vector. This way, we query block assignments from compression vectors with a fewer number of segments, thereby getting faster access times. We demonstrate the speed up offered by this approach \hbox{in Section \ref{subsubsec:batch-wise_exploration}.}

\subsection{Fennel Modification}
\label{subsec:kappa_modification}
As mentioned previously, the efficiency of on-the-fly compression of block assignments is greatly affected by whether the partitioner’s block assignments create long or short runs. We thus motivate our partitioner to generate long runs by incentivizing it to place the currently streamed node into the block assigned to the previously streamed node. To do so, we modify the score generated by the \textsc{Fennel} objective function by introducing a scaling parameter $\kappa$ \hbox{as follows:}

Let $b_{\text{prev}}$ be the block assigned to the previously streamed node. Then, after computing the \textsc{Fennel} score $S$ for block $b$ for the current node, get an updated score $S_{\text{new}}$ by calling
\[
S_{\text{new}} = f(b, S) = 
\begin{cases} 
	S \cdot \left( \kappa^{\text{sgn}(S)} \right) & \text{if } b = b_{\text{prev}} \\
	S & \text{if } b \neq b_{\text{prev}}
\end{cases}
\]

The function $f(b, S)$ takes two parameters: $b$, which refers to the partition block, and $S$, which refers to the score produced by the \textsc{Fennel} objective function for block $b$. If block $b$ is the same as the previous block $b_{\text{prev}}$, then the score $S$ is multiplied by $\kappa^{\text{sgn}(x)}$. Here, $\text{sgn}(x)$ returns $1$ for positive $x$, $-1$ for negative $x$, and $0$ for $x = 0$. So, the function effectively multiplies positive scores by $\kappa$ and divides negative scores by $\kappa$. If block $b$ is different from the previous block $b_{\text{prev}}$, then \hbox{the score~$S$ remains unchanged.}

The $\kappa$ parameter ensures that the score returned by the \textsc{Fennel} objective function is scaled by a factor of $\kappa$ for the previously assigned block, even if the score produced a negative value. As a consequence, the partitioner is more likely to select the same block for the current node as for the previous one, thereby producing longer runs of block assignments. In Section~\ref{subsubsec:fennel_mod}, we examine the performance of the partitioner with varying values of $\kappa$. 

\subsection{External Memory Priority Queue}
\label{subsec:expq}
As an alternative to using run-length compression to reduce the memory requirements of the streaming partitioner, we implement an external memory priority queue to store block assignments on the disk. Since we wish to access blocks assigned to neighbors of the currently streamed node, we use time-forward processing. A pre-requisite of time-forward processing is that the directed acyclic graph (DAG) corresponding to the data must be stored in a topological order. Thus, we model our graph as a DAG by considering only edges $(v,w)$ where $v < w$. After assigning a block $b$ to the currently streamed node $v$, we insert a pair $\langle\text{block ID}, \text{target node}\rangle$ for every neighbor $w$ of $v$, i.e., we insert \hbox{$\langle b, w \rangle\ \forall w \in N(v) \mid v < w$} into the priority queue. The target node in the pair serves as the priority with which to build the priority queue, while the block ID is the value we wish to extract when accessing and deleting the minimum element of the priority queue. Then, when we are streaming a future node $u$, we call \texttt{extractMin(PQ)} $\ell$ times where $\ell$ is the number of neighbors $w$ of $u$ such that $w < u$. This fetches all block assignments that are relevant for computing the partitioning scoring function for the current node. In our implementation, we use the \textsc{stxxl} \hbox{external memory priority queue.} The \textsc{stxxl} priority queue is maintained using the sequence heap operation, which takes $\mathcal{O}(\frac{1}{M_B}\log_{M/M_B}(I/M_B))$ amortized I/Os, where $I$ is the number of insertions and $M$ is the size of internal memory that can access external memory by transferring blocks of size $M_B$~\cite{SandersSTXXL}. Both insertion and deletion require $\mathcal{O}(\log I)$ work and $\mathcal{O}(1/M_B)$ amortized I/Os. Here, we have $I = m/2$ insertions and deletions, as we insert once for every neighbor of the current node with a larger node ID, and delete once for every neighbor of the current node with \hbox{a smaller node ID}. 

\section{Experimental Evaluation}
\label{sec:Experimental Evaluation}

\textbf{Setup.}
We implemented the (semi-)dynamic run-length compression vector as a standalone templated container in \textsc{C++}. 
We use this compression vector in \framework, along with a modified \textsc{Fennel} scoring function which drops runtime dependency on $k$~\cite{freight_paper}. Any scoring function, like \textsc{LDG} could be used in place of \textsc{Fennel}, and \framework\ can be plugged into any buffered streaming partitioner. 
\framework\ is compiled using gcc 13.2.0 with full optimization enabled (-O3 flag).
All parameter study experiments were performed on a single core of a machine with a sixteen-core Intel Xeon Silver 4216 processor running at $2.1$ GHz, $100$ GB of main memory, $16$ MB of L2-Cache, and $22$ MB of L3-Cache running Ubuntu 20.04.1. %
To demonstrate the extent of the capabilities of our framework, we additionally perform experiments on a Raspberry Pi 5 with a Broadcom BCM2712 2.4 GHz quad-core 64-bit Arm Cortex-A76 CPU \hbox{and 8GB of LPDDR4X-4267 SDRAM.}

\textbf{Instances.}
\begin{table*}[t]
	\centering
	\caption{
		Benchmark graphs used in our experiments. 
		\instance{roadNet-*}, \instance{wiki-*}, \instance{web-*}, all social, co-purchasing, and autonomous systems graphs were obtained from the publicly available SNAP dataset~\cite{snap}. 
		\instance{eu-2005}, \instance{in-2004} and \instance{uk-2007-05} were obtained from the 10th DIMACS Implementation Challenge benchmark set~\cite{benchmarksfornetworksanalysis}.
		All remaining graphs are available on the network repository website~\cite{nr-aaai15} or on the Laboratory for Web Algorithmics website~\cite{BMSB,BRSLLP,BoVWFI}, including \texttt{gsh-2015}, \texttt{clueweb12}, and \texttt{uk-2014}, which are some of the largest publicly available graph networks. \texttt{RGG} and \texttt{BA} are randomly generated graphs of type \texttt{Random Geometric Graph} and \texttt{Barabassi-Albert} respectively~\cite{KaGen,BAGenerator}. 
		For our experiments, we converted these graphs to a node-stream format (METIS) while removing parallel edges, self-loops, and directions, and assigning unitary weight to all nodes and edges.
	} 
	\setlength{\tabcolsep}{3.7pt}
	\begin{tabular}{lrrrclrrr}
		\textsc{Graph} & $n$ & $m$ & \textsc{Type} & & \textsc{Graph} & $n$ & $m$ & \textsc{Type} \\
		\midrule
		\benchSet{Exploration \& Tuning Set} \\
		\midrule
		\instanceSmall{Dubcova1} & 16\,129 & 118\,440 & \instMesh && \instanceSmall{roadNet-TX} & 1\,379\,917 & 1\,921\,660 & \instRoad \\
		\instanceSmall{DBLP-2010} & 300\,647 & 807\,700 & \instCitation && \instanceSmall{in-2004} & 1\,382\,908 & 13\,591\,473 & \instWeb \\
		\instanceSmall{com-DBLP} & 317\,080 & 1\,049\,866 & \instSocial && \instanceSmall{Flan\_1565} & 1\,564\,794 & 57\,920\,625 & \instMesh \\
		\instanceSmall{web-NotreDame} & 325\,729 & 1\,090\,108 & \instWeb && \instanceSmall{G3\_Circuit} & 1\,585\,478 & 3\,037\,674 & \instCircuit \\
		\instanceSmall{com-Amazon} & 334\,863 & 925\,872 & \instSocial && \instanceSmall{soc-pokec} & 1\,632\,803 & 22\,301\,964 & \instSocial \\
		\instanceSmall{web-Google} & 356\,648 & 2\,093\,324 & \instWeb && \instanceSmall{wiki-topcats} & 1\,791\,489 & 25\,444\,207 & \instSocial \\
		\instanceSmall{ML\_Laplace} & 377\,002 & 13\,656\,485 & \instMesh && \instanceSmall{roadNet-CA} & 1\,965\,206 & 2\,766\,607 & \instRoad \\
		\instanceSmall{amazon0312} & 400\,727 & 2\,349\,869 & \instCoPurch && \instanceSmall{HV15R} & 2\,017\,169 & 162\,357\,569 & \instMesh \\
		\instanceSmall{amazon0601} & 403\,394 & 2\,443\,408 & \instCoPurch && \instanceSmall{soc-flixster} & 2\,523\,386 & 7\,918\,801 & \instSocial \\
		\instanceSmall{amazon0505} & 410\,236 & 2\,439\,437 & \instCoPurch && \instanceSmall{Bump\_2911} & 2\,852\,430 & 62\,409\,240 & \instMesh \\
		\instanceSmall{coPapersCiteseer} & 434\,102 & 16\,036\,720 & \instCitation && \instanceSmall{FullChip} & 2\,986\,999 & 11\,817\,567 & \instCircuit \\
		\instanceSmall{coPapersDBLP} & 540\,486 & 15\,245\,729 & \instCitation && \instanceSmall{com-Orkut} & 3\,072\,441 & 117\,185\,083 & \instSocial \\
		\instanceSmall{as-Skitter} & 554\,930 & 5\,797\,633 & \instAutSyst && \instanceSmall{cit-Patents} & 3\,774\,768 & 16\,518\,947 & \instCitation \\
		\instanceSmall{Amazon-2008} & 735\,323 & 3\,523\,472 & \instSimilarity && \instanceSmall{com-LJ} & 3\,997\,962 & 34\,681\,189 & \instSocial \\
		\instanceSmall{eu-2005} & 862\,664 & 16\,138\,468 & \instWeb && \instanceSmall{soc-LiveJournal1} & 4\,846\,609 & 42\,851\,237 & \instSocial \\
		\instanceSmall{ca-hollywood-2009} & 1\,069\,126 & 56\,306\,653 & \instRoad && \instanceSmall{Ljournal-2008} & 5\,363\,186 & 49\,514\,271 & \instSocial \\
		\instanceSmall{roadNet-PA} & 1\,088\,092 & 1\,541\,898 & \instRoad && \instanceSmall{circuit5M} & 5\,558\,311 & 26\,983\,926 & \instCircuit \\
		\instanceSmall{com-Youtube} & 1\,134\,890 & 2\,987\,624 & \instSocial && \instanceSmall{italy-osm} & 6\,686\,493 & 7\,013\,978 & \instRoad \\
		\instanceSmall{soc-lastfm} & 1\,191\,805 & 4\,519\,330 & \instSocial && \instanceSmall{great-britain-osm} & 7\,733\,822 & 8\,156\,517 & \instRoad \\
		\midrule
		\benchSet{Test Set} \\
		\midrule
		\instanceSmall{orkut} & 3\,072\,411 & 117\,185\,082 & \instSocial && \instanceSmall{com-Friendster} & 65\,608\,366 & 1\,806\,067\,135 & \instSocial \\
		\instanceSmall{arabic-2005} & 22\,744\,080 & 553\,903\,073 & \instWeb && \instanceSmall{rgg26} & 67\,108\,864 & 574\,553\,645 & \instGen \\
		\instanceSmall{nlpkkt240} & 27\,933\,600 & 373\,239\,376 & \instMatrix && \instanceSmall{rhg1B} & 100\,000\,000 & 1\,000\,913\,106 & \instGen \\
		\instanceSmall{it-2004} & 41\,291\,594 & 1\,027\,474\,947 & \instWeb && \instanceSmall{rhg2B} & 100\,000\,000 & 1\,999\,544\,833 & \instGen \\
		\instanceSmall{twitter-2010} & 41\,652\,230 & 1\,202\,513\,046 & \instSocial && \instanceSmall{uk-2007-05} & 105\,896\,555 & 3\,301\,876\,564 & \instWeb \\
		\instanceSmall{sk-2005} & 50\,636\,154 & 1\,810\,063\,330 & \instWeb && \instanceSmall{webbase-2001} & 118\,142\,155 & 854\,809\,761 & \instWeb \\
		\midrule
		\benchSet{Massive Set} \\
		\midrule
		\instanceSmall{gsh-2015} & \small{988\,490\,691} & \small{25\,690\,705\,118} & \instWeb && \instanceSmall{BA} & \small{17\,179\,869\,184} & \small{1\,030\,792\,001\,284} & \instGen \\
		\instanceSmall{clueweb12} & \small{978\,408\,098} & \small{37\,372\,179\,311} & \instWeb && \instanceSmall{RGG} & \small{17\,179\,869\,184} & \small{1\,031\,585\,082\,692} & \instGen \\
		\instanceSmall{uk-2014} & \small{787\,801\,471} & \small{42\,464\,215\,550} & \instWeb && & & & \\
		\bottomrule
		\vspace{1pt}
	\end{tabular}
	\label{tab:instances}
\end{table*}
Our graph instances for experiments are shown in Table~\ref{tab:instances} and are sourced from Ref.~\cite{benchmarksfornetworksanalysis,BMSB,BRSLLP,BoVWFI,KaGen,snap,nr-aaai15}.
All instances evaluated have been used for benchmarking in previous works on graph partitioning.
We set the number of blocks to $k = \{2^1, 2^2, \dots, 2^{8}\}$ for tuning and test experiments.
We allow an imbalance of $\varepsilon = 3\%$ for all partitioners.
While streaming, we use the default order of nodes in these graphs, i.e., we do not change the ordering of nodes in the sourced instances. 

\textbf{Streaming Graph Generation.}
To evaluate our framework on large-scale graphs, we developed a streaming version of the \textsc{KaGen} graph generator~\cite{KaGen,BAGenerator}, which supports a variety of graph generation models, including random geometric, random Delaunay, random hyperbolic, and Barabási-Albert graphs. This streaming generator mitigates the I/O constraints typical of resource-limited platforms like the Raspberry Pi and will be made available as open-source software for users to experiment with streaming graph algorithms across different applications.

Our streaming approach allows flexible selection of both the graph model and the construction parameters, such as the number of nodes or edges. The generator sequentially returns either an edge list or the neighborhood of each successive node, adhering to common graph storage formats. Internally, the graph is constructed incrementally in chunks, meaning only a small portion of the graph resides in memory at any given time. Here, on-the-fly graph generation is made possible using a communication-free paradigm that involves re-computation with psuedorandom hashing and various divide-and-conquer schemes, as detailed in ~\cite{KaGen,BAGenerator}. The number of chunks is controlled by the user, providing a trade-off between runtime and memory consumption. 

In this work, we conduct experiments on randomly generated billion-node, trillion-edge graphs using the Barabási-Albert (\texttt{BA}) and random geometric (\texttt{RGG}) models to represent large scale-free networks and mesh-like graphs, respectively. We set the average degree to 60 for the \texttt{BA} graphs and use a radius of 4.71536 $\times$ $10^{-5}$ for the \texttt{RGG} graphs. These parameters, which are within the typical ranges observed in real-world networks, allow us to generate graphs with a comparable number of edges across both models.

\textbf{Methodology.}
We measure running time, edge cut, and memory consumption, i.e., the maximum resident set size for the executed process.
When averaging over all instances, we use the geometric mean to give every instance the same influence on the final score.
Let the runtime, edge cut, or memory consumption be denoted by the score $\sigma_{A}$ for some $k$-partition generated by an algorithm $A$.
We express this score relative to others using the following tools:
\emph{improvement} over an algorithm $B$, computed as a percentage $(\frac{\sigma_A}{\sigma_B} - 1) * 100 \%$ and
\emph{relative} value over an algorithm $B$, computed as $\frac{\sigma_A}{\sigma_B}$.
Additionally, we present performance profiles by Dolan and Mor{\'e}~\cite{pp} to benchmark our algorithms.
These profiles relate the running time (resp. solution quality, memory) of the slower (resp. worse) algorithms to the fastest (resp. best) one on a per-instance basis, rather than grouped by $k$.
Their $x$-axis shows a factor $\tau$ while their $y$-axis shows the fraction of instances for which an algorithm has up to $\tau$ times the running time (resp. solution quality, memory) of the fastest \hbox{(resp. best) algorithm}.

\subsection{Parameter Study}
We begin by exploring possible variations to the parameters used in \framework, with different memory-runtime trade-offs, as described in Section~\ref{sec:main_contribution}, namely the use of batch-wise compression to increase append speed and a modified \textsc{Fennel} scoring function scaled by $\kappa$ to improve on-the-fly compression. We present the results of experiments that explored the performance of our streaming partitioner using different batch sizes $\beta$ for the compression vector (Section~\ref{subsubsec:batch-wise_exploration}) and different $\kappa$ values for \textsc{Fennel} modification (Section~\ref{subsubsec:fennel_mod}). We compare these to the use of default \framework, which is not batch-wise, i.e, $\beta = n$ and not $\kappa$-modified, i.e, $\kappa = 1$. Hereafter, we explicitly refer to these parameters with subscripts, such as $\defaultframework$.

\subsubsection{Batch-Wise Mode Exploration}
\label{subsubsec:batch-wise_exploration}
\begin{figure}[t]
	\centering
	\input{plots/perf_batch}
	\begin{tikzpicture}[x=1pt,y=1pt]
\definecolor{fillColor}{RGB}{255,255,255}
\begin{scope}
\definecolor{fillColor}{RGB}{255,255,255}

\path[fill=fillColor] (-20.77,250.10) rectangle ( -3.42,267.44);
\end{scope}
\begin{scope}
\definecolor{drawColor}{RGB}{77,175,74}

\path[draw=drawColor,line width= 1.7pt,line join=round] (-19.03,258.77) -- ( -5.16,258.77);
\end{scope}
\begin{scope}
\definecolor{fillColor}{RGB}{77,175,74}

\path[fill=fillColor] (-14.27,256.60) --
	( -9.92,256.60) --
	( -9.92,260.95) --
	(-14.27,260.95) --
	cycle;
\end{scope}
\begin{scope}
\definecolor{fillColor}{RGB}{255,255,255}

\path[fill=fillColor] (-20.77,238.45) rectangle ( -3.42,255.79);
\end{scope}
\begin{scope}
\definecolor{drawColor}{RGB}{55,126,184}

\path[draw=drawColor,line width= 1.7pt,dash pattern=on 2pt off 2pt ,line join=round] (-19.03,247.12) -- ( -5.16,247.12);
\end{scope}
\begin{scope}
\definecolor{fillColor}{RGB}{55,126,184}

\path[fill=fillColor] (-12.09,247.12) circle (  2.18);
\end{scope}
\begin{scope}
\definecolor{fillColor}{RGB}{255,255,255}

\path[fill=fillColor] ( 88.68,250.10) rectangle (106.02,267.44);
\end{scope}
\begin{scope}
\definecolor{drawColor}{RGB}{228,26,28}

\path[draw=drawColor,line width= 1.7pt,dash pattern=on 4pt off 2pt ,line join=round] ( 90.41,258.77) -- (104.29,258.77);
\end{scope}
\begin{scope}
\definecolor{fillColor}{RGB}{228,26,28}

\path[fill=fillColor] ( 97.35,262.16) --
	(100.28,257.08) --
	( 94.42,257.08) --
	cycle;
\end{scope}
\begin{scope}
\definecolor{drawColor}{RGB}{0,0,0}

\node[text=drawColor,anchor=base west,inner sep=0pt, outer sep=0pt, scale=  0.80] at (  1.58,256.02) {\textsc{StreamCpi}$_{\beta=n, \kappa=1}$};
\end{scope}
\begin{scope}
\definecolor{drawColor}{RGB}{0,0,0}

\node[text=drawColor,anchor=base west,inner sep=0pt, outer sep=0pt, scale=  0.80] at (  1.58,244.36) {\textsc{StreamCpi}$_{\beta=10k, \kappa=1}$};
\end{scope}
\begin{scope}
\definecolor{drawColor}{RGB}{0,0,0}

\node[text=drawColor,anchor=base west,inner sep=0pt, outer sep=0pt, scale=  0.80] at (111.02,256.02) {\textsc{StreamCpi}$_{\beta=100k, \kappa=1}$};
\end{scope}
\end{tikzpicture}
	\caption{Comparison of $\defaultframework$ with $\framework_{\beta=10k,\kappa=1}$ and $\framework_{\beta=100k,\kappa=1}$, i.e., \textit{batch-wise} \framework, with a batch size of 10\,000 and 100\,000, on the Tuning Set using performance profiles.}
	\label{plot:batch-wise}
\end{figure}
\begin{figure*}[t]
	\centering
	\input{plots/perf_kappa}
	\begin{tikzpicture}[x=1pt,y=1pt]
\definecolor{fillColor}{RGB}{255,255,255}
\begin{scope}
\definecolor{fillColor}{RGB}{255,255,255}

\path[fill=fillColor] (-93.70,250.10) rectangle (-76.35,267.44);
\end{scope}
\begin{scope}
\definecolor{drawColor}{RGB}{77,175,74}

\path[draw=drawColor,line width= 1.7pt,line join=round] (-91.96,258.77) -- (-78.09,258.77);
\end{scope}
\begin{scope}
\definecolor{fillColor}{RGB}{77,175,74}

\path[fill=fillColor] (-87.20,256.60) --
	(-82.85,256.60) --
	(-82.85,260.95) --
	(-87.20,260.95) --
	cycle;
\end{scope}
\begin{scope}
\definecolor{fillColor}{RGB}{255,255,255}

\path[fill=fillColor] (-93.70,238.45) rectangle (-76.35,255.79);
\end{scope}
\begin{scope}
\definecolor{drawColor}{RGB}{152,78,163}

\path[draw=drawColor,line width= 1.7pt,dash pattern=on 2pt off 2pt ,line join=round] (-91.96,247.12) -- (-78.09,247.12);
\end{scope}
\begin{scope}
\definecolor{fillColor}{RGB}{152,78,163}

\path[fill=fillColor] (-85.02,247.12) circle (  2.18);
\end{scope}
\begin{scope}
\definecolor{fillColor}{RGB}{255,255,255}

\path[fill=fillColor] ( 35.88,250.10) rectangle ( 53.23,267.44);
\end{scope}
\begin{scope}
\definecolor{drawColor}{RGB}{255,127,0}

\path[draw=drawColor,line width= 1.7pt,dash pattern=on 4pt off 2pt ,line join=round] ( 37.62,258.77) -- ( 51.49,258.77);
\end{scope}
\begin{scope}
\definecolor{fillColor}{RGB}{255,127,0}

\path[fill=fillColor] ( 44.56,262.16) --
	( 47.49,257.08) --
	( 41.62,257.08) --
	cycle;
\end{scope}
\begin{scope}
\definecolor{fillColor}{RGB}{255,255,255}

\path[fill=fillColor] ( 35.88,238.45) rectangle ( 53.23,255.79);
\end{scope}
\begin{scope}
\definecolor{drawColor}{RGB}{166,86,40}

\path[draw=drawColor,line width= 1.7pt,dash pattern=on 4pt off 4pt ,line join=round] ( 37.62,247.12) -- ( 51.49,247.12);
\end{scope}
\begin{scope}
\definecolor{fillColor}{RGB}{166,86,40}

\path[fill=fillColor] ( 42.38,247.12) --
	( 44.56,249.29) --
	( 46.73,247.12) --
	( 44.56,244.94) --
	cycle;
\end{scope}
\begin{scope}
\definecolor{fillColor}{RGB}{255,255,255}

\path[fill=fillColor] (169.92,250.10) rectangle (187.27,267.44);
\end{scope}
\begin{scope}
\definecolor{drawColor}{RGB}{247,129,191}

\path[draw=drawColor,line width= 1.7pt,dash pattern=on 1pt off 3pt ,line join=round] (171.66,258.77) -- (185.53,258.77);
\end{scope}
\begin{scope}
\definecolor{drawColor}{RGB}{247,129,191}

\path[draw=drawColor,line width= 0.4pt,line join=round,line cap=round] (178.59,255.39) --
	(181.53,260.46) --
	(175.66,260.46) --
	cycle;
\end{scope}
\begin{scope}
\definecolor{drawColor}{RGB}{0,0,0}

\node[text=drawColor,anchor=base west,inner sep=0pt, outer sep=0pt, scale=  0.80] at (-71.35,256.02) {\textsc{StreamCpi}$_{\beta=n, \kappa=1}$};
\end{scope}
\begin{scope}
\definecolor{drawColor}{RGB}{0,0,0}

\node[text=drawColor,anchor=base west,inner sep=0pt, outer sep=0pt, scale=  0.80] at (-71.35,244.36) {\textsc{StreamCpi}$_{\beta=n, \kappa=5}$};
\end{scope}
\begin{scope}
\definecolor{drawColor}{RGB}{0,0,0}

\node[text=drawColor,anchor=base west,inner sep=0pt, outer sep=0pt, scale=  0.80] at ( 58.23,256.02) {\textsc{StreamCpi}$_{\beta=n, \kappa=10}$};
\end{scope}
\begin{scope}
\definecolor{drawColor}{RGB}{0,0,0}

\node[text=drawColor,anchor=base west,inner sep=0pt, outer sep=0pt, scale=  0.80] at ( 58.23,244.36) {\textsc{StreamCpi}$_{\beta=n, \kappa=15}$};
\end{scope}
\begin{scope}
\definecolor{drawColor}{RGB}{0,0,0}

\node[text=drawColor,anchor=base west,inner sep=0pt, outer sep=0pt, scale=  0.80] at (192.27,256.02) {\textsc{StreamCpi}$_{\beta=n, \kappa=20}$};
\end{scope}
\end{tikzpicture}
	\caption{Comparison of $\defaultframework$ with $\framework_{\beta=n,\kappa=5}$, $\framework_{\beta=n,\kappa=10}$, $\framework_{\beta=n,\kappa=15}$, $\framework_{\beta=n,\kappa=20}$, i.e., $\kappa$\textit{-modified} \framework, with $\kappa$ = 5, 10, 15, and 20, on the Tuning Set using performance profiles.}
	\label{plot:kappa}
\end{figure*}

\begin{figure*}[ht]
	\centering
	\input{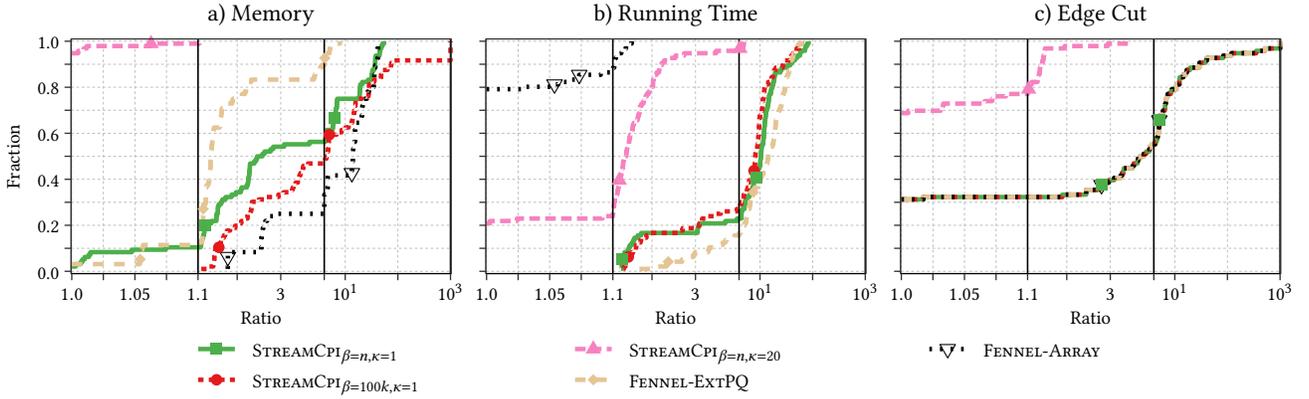}
	\begin{tikzpicture}[x=1pt,y=1pt]
\definecolor{fillColor}{RGB}{255,255,255}
\begin{scope}
\definecolor{fillColor}{RGB}{255,255,255}

\path[fill=fillColor] (-60.45,250.10) rectangle (-43.11,267.44);
\end{scope}
\begin{scope}
\definecolor{drawColor}{RGB}{77,175,74}

\path[draw=drawColor,line width= 1.7pt,line join=round] (-58.72,258.77) -- (-44.84,258.77);
\end{scope}
\begin{scope}
\definecolor{fillColor}{RGB}{77,175,74}

\path[fill=fillColor] (-53.96,256.60) --
	(-49.61,256.60) --
	(-49.61,260.95) --
	(-53.96,260.95) --
	cycle;
\end{scope}
\begin{scope}
\definecolor{fillColor}{RGB}{255,255,255}

\path[fill=fillColor] (-60.45,238.45) rectangle (-43.11,255.79);
\end{scope}
\begin{scope}
\definecolor{drawColor}{RGB}{228,26,28}

\path[draw=drawColor,line width= 1.7pt,dash pattern=on 2pt off 2pt ,line join=round] (-58.72,247.12) -- (-44.84,247.12);
\end{scope}
\begin{scope}
\definecolor{fillColor}{RGB}{228,26,28}

\path[fill=fillColor] (-51.78,247.12) circle (  2.18);
\end{scope}
\begin{scope}
\definecolor{fillColor}{RGB}{255,255,255}

\path[fill=fillColor] ( 82.06,250.10) rectangle ( 99.41,267.44);
\end{scope}
\begin{scope}
\definecolor{drawColor}{RGB}{247,129,191}

\path[draw=drawColor,line width= 1.7pt,dash pattern=on 4pt off 2pt ,line join=round] ( 83.80,258.77) -- ( 97.67,258.77);
\end{scope}
\begin{scope}
\definecolor{fillColor}{RGB}{247,129,191}

\path[fill=fillColor] ( 90.74,262.16) --
	( 93.67,257.08) --
	( 87.80,257.08) --
	cycle;
\end{scope}
\begin{scope}
\definecolor{fillColor}{RGB}{255,255,255}

\path[fill=fillColor] ( 82.06,238.45) rectangle ( 99.41,255.79);
\end{scope}
\begin{scope}
\definecolor{drawColor}{RGB}{229,196,148}

\path[draw=drawColor,line width= 1.7pt,dash pattern=on 4pt off 4pt ,line join=round] ( 83.80,247.12) -- ( 97.67,247.12);
\end{scope}
\begin{scope}
\definecolor{fillColor}{RGB}{229,196,148}

\path[fill=fillColor] ( 88.56,247.12) --
	( 90.74,249.29) --
	( 92.91,247.12) --
	( 90.74,244.94) --
	cycle;
\end{scope}
\begin{scope}
\definecolor{fillColor}{RGB}{255,255,255}

\path[fill=fillColor] (216.10,250.10) rectangle (233.45,267.44);
\end{scope}
\begin{scope}
\definecolor{drawColor}{RGB}{0,0,0}

\path[draw=drawColor,line width= 1.7pt,dash pattern=on 1pt off 3pt ,line join=round] (217.84,258.77) -- (231.71,258.77);
\end{scope}
\begin{scope}
\definecolor{drawColor}{RGB}{0,0,0}

\path[draw=drawColor,line width= 0.4pt,line join=round,line cap=round] (224.78,255.39) --
	(227.71,260.46) --
	(221.84,260.46) --
	cycle;
\end{scope}
\begin{scope}
\definecolor{drawColor}{RGB}{0,0,0}

\node[text=drawColor,anchor=base west,inner sep=0pt, outer sep=0pt, scale=  0.80] at (-38.11,256.02) {\textsc{StreamCpi}$_{\beta=n, \kappa=1}$};
\end{scope}
\begin{scope}
\definecolor{drawColor}{RGB}{0,0,0}

\node[text=drawColor,anchor=base west,inner sep=0pt, outer sep=0pt, scale=  0.80] at (-38.11,244.36) {\textsc{StreamCpi}$_{\beta=100k, \kappa=1}$};
\end{scope}
\begin{scope}
\definecolor{drawColor}{RGB}{0,0,0}

\node[text=drawColor,anchor=base west,inner sep=0pt, outer sep=0pt, scale=  0.80] at (104.41,256.02) {\textsc{StreamCpi}$_{\beta=n, \kappa=20}$};
\end{scope}
\begin{scope}
\definecolor{drawColor}{RGB}{0,0,0}

\node[text=drawColor,anchor=base west,inner sep=0pt, outer sep=0pt, scale=  0.80] at (104.41,244.36) {\textsc{Fennel-ExtPQ}};
\end{scope}
\begin{scope}
\definecolor{drawColor}{RGB}{0,0,0}

\node[text=drawColor,anchor=base west,inner sep=0pt, outer sep=0pt, scale=  0.80] at (238.45,256.02) {\textsc{Fennel-Array}};
\end{scope}
\end{tikzpicture}
	\caption{Comparison of default \framework, with \textit{batch-wise} and $\kappa$\textit{-modified} \framework, \textsc{Fennel-Array} and \textsc{Fennel-ExtPQ} on the Test Set using performance profiles.}
	\label{plot:test}
\end{figure*}

Our experiments indicate that batch-wise compression (Section~\ref{subsec:batch_cpi}) decreases overall runtime but increases memory overhead compared to $\defaultframework$, with time efficiency decreasing and memory efficiency increasing, respectively, with increasing batch size relative to the size of the graph. This result is intuitive, as the use of batches increases the number of compression vectors, thus requiring additional memory, but enables faster append as fewer runs are appended to shorter bit vectors. The results from our Tuning Set indicate that, on average, $\framework_{\beta=10k,\kappa=1}$ was $2.9\times$ faster than $\defaultframework$, but consumed $1.3\times$ more memory. Using $\framework_{\beta=100k,\kappa=1}$, produced, on average, a slower speedup of $1.15\times$ over \linebreak $\defaultframework$, while consuming $1.1\times$ more memory. Figure~\ref{plot:batch-wise} indicates that the use of $\defaultframework$ consumes the least memory in almost 80\% of all instances; the batch-size of \linebreak $\framework_{\beta=100k,\kappa=1}$ consumes less memory than the smaller batch size of $\framework_{\beta=10k,\kappa=1}$ in all instances. Figure~\ref{plot:batch-wise} shows that, in contrast to memory consumption, the use of a smaller batch-size of $\framework_{\beta=10k,\kappa=1}$ produces the fastest runtime in over 95\% of all instances, with $\defaultframework$ having the slowest runtime in almost all instances.

\subsubsection{Fennel Modification.}
\label{subsubsec:fennel_mod}
Modifying the \textsc{Fennel} scoring function with a $\kappa$ scaling parameter, as described in Section~\ref{subsec:kappa_modification}, reduces both memory and runtime overhead, with our experiments indicating that $\kappa=20$ performs best. Experiments on the Tuning Set demonstrate that, on average, $\framework_{\beta=n,\kappa=5}$ is $5.4\times$ faster, and uses $1.6\times$ less memory than $\defaultframework$. With $\framework_{\beta=n,\kappa=20}$, the partitioner is $6.3\times$ faster and uses $1.7\times$ less memory on average, compared to $\defaultframework$. 
We determine $\kappa=20$ to be a suitable $\kappa$-value for memory and runtime efficiency, as further increases to $\kappa$ yield comparable or worse memory, runtime and solution quality trade-offs. 
Figure \ref{plot:kappa} demonstrates that using $\kappa$-modification produces lower memory overhead and faster runtime compared to $\defaultframework$ in all instances. However, our results show that $\kappa$\textit{-modified} \framework\ in Tuning Set instances produces worse solution quality than $\defaultframework$; on average across all instances, \framework\ with $\kappa=1$ produces between $7.7\%$ to $10\%$ better solution quality than \framework\ with $\kappa = 5$ up to $\kappa = 20$. Note, however, as Figure~\ref{plot:kappa} demonstrates, $\kappa$\textit{-modified} \framework\ produces better solution quality than $\defaultframework$ in approximately 10\% of instances. In these instances, nodes are streamed based on some notion of high locality, i.e., nodes are ordered close to \hbox{their neighbors.} 

\emph{In summary}, our parameter study offers two categories of modes for \framework: the batch-wise compression mode to optimize runtime, at the cost of memory overhead, and the $\kappa$\textit{-modified} \framework\ to optimize both runtime and memory, at the potential cost \hbox{of solution quality.}

\subsection{Comparison with Standard Array, External Memory Priority Queue and Hashing}
We present experimental results from the Test Set depicting the relative performance of \framework\ using the different modes of our compression vector -- default $\defaultframework$, \textit{batch-wise} $\framework_{\beta=100k,\kappa=1}$, and $\kappa$\textit{-modified} $\framework_{\beta=n,\kappa=20}$ -- as compared to \textsc{Fennel} with a standard array (hereafter ‘\textsc{Fennel-Array}’), and using the external memory priority queue (hereafter ‘\textsc{Fennel-ExtPQ}’) that we presented in Section~\ref{subsec:expq}. 
Here, we exclude results for \framework\ using $\kappa$-modification in combination with \textit{batch-wise} compression, as the overall compression achieved by setting $\kappa=20$ is high enough that \textit{batch-wise} compression does not offer additional benefits. 
Figure \ref{plot:test} gives performance profiles for the Test Set. We find that the use of the $\kappa$ modification offers the best memory optimization, without significantly increasing runtime: our results show that, on average, $\framework_{\beta=n,\kappa=20}$ consumes $9.7\times$ less memory than \textsc{Fennel-Array}, while being $1.5\times$ slower. $\framework_{\beta=n,\kappa=20}$ uses the least memory in almost all Test instances, with a comparable runtime to \textsc{Fennel-Array} (Figure \ref{plot:test}). Surprisingly, due to high locality in the graph instances,\linebreak $\framework_{\beta=n,\kappa=20}$ produces 76.44\% better solution quality on average than the rest, which all have the same solution quality. 

Our experimental results show that, on average, our proposed \framework\ framework with tunable compression modes and external memory priority queue all reduce memory overhead, as compared to \textsc{Fennel-Array}. On average, $\defaultframework$ uses \linebreak 
$2.2\times$ less memory than \textsc{Fennel-Array}, at the cost of being $8.2\times$ slower. The \textit{batch-wise} $\framework_{\beta=100k,\kappa=1}$ improves on \linebreak $\defaultframework$ in runtime, being $6.9\times$ slower than \linebreak \textsc{Fennel-Array} on average, but uses more memory ($2.2\times$ more memory on average than \textsc{Fennel-Array}). As shown in Figure~\ref{plot:test}, both $\defaultframework$ and
$\framework_{\beta=100k,\kappa=1}$ consume less memory than \textsc{Fennel-Array} for over 80\% of all instances, but have a longer runtime in all instances. Those instances in which \textsc{Fennel-Array} uses less memory than \framework\ represent graph partitions with many short runs of consecutive block assignments, for which compression is less effective. The use of an external memory priority queue produces significant memory improvements, using $5.8\times$ less memory than \textsc{Fennel-Array} on average, but has a higher runtime cost, as it is on average $12.0\times$ slower than \textsc{Fennel-Array}. Additionally, it uses less memory than $\defaultframework$ in all instances, but is slower than it in about 90\% of instances (Figure~\ref{plot:test}). This is because the \textsc{stxxl} external memory priority queue has been configured to use as little data in the internal memory as possible (internal buffer size of 3MB), reducing memory consumption but increasing runtime due to more frequent I/O from disk. One advantage of \textsc{Fennel-ExtPQ} is that its memory consumption is arbitrarily configurable by changing the allowed size of the internal memory buffer, but this is a trade-off with running time. Nonetheless, memory-optimized \textsc{Fennel-ExtPQ} still consumes $1.7\times$ more memory than $\framework_{\beta=n,\kappa=20}$ on average. 

\emph{In summary}, all the \framework\ compression modes and external memory priority queue that we propose reduce the memory consumption of \textsc{Fennel-Array}, i.e., the \textsc{Fennel} partitioner using a standard array to store block assignments. Employing $\kappa$ modification, i.e., $\framework_{\beta=n,\kappa=20}$, produces the best compression, resulting in, on average, memory overhead that is $9.7\times$ less than \textsc{Fennel-Array}, while being $1.5\times$ slower. Using the $\kappa$ modified $\framework_{\beta=n,\kappa=20}$ additionally improves solution quality by 76.44\% on average, as it produces significant solution quality improvements in high-locality graphs (though it would reduce solution quality in low-locality graphs).
Our \textit{default} $\defaultframework$ and \textit{batch-wise} $\framework_{\beta=100k,\kappa=1}$, use, on average, $2.2\times$ less memory and $2.2\times$ more memory, respectively, than the standard array, and are $8.2\times$ and $6.9\times$ slower respectively. The external memory priority queue reduces memory significantly, using $5.8\times$ less memory on average than the standard array, but is $12.0\times$ slower.

\subsection{Partitioning on Raspberry Pi}
\label{subsec:rasp_pi}
\newcommand{\ccHead}[1]{\multicolumn{1}{c}{#1}}
\newcommand{\capCut}{\textsc{RelCut}}
\newcommand{\capTime}{\textsc{Rt} [s]}
\newcommand{\capMemory}{\textsc{Mem} [GB]}
\newcommand{\headCut}{\ccHead{\capCut}}
\newcommand{\headTime}{\ccHead{\capTime}}
\newcommand{\headMemory}{\ccHead{\capMemory}}
\newcommand*{\xdash}[1][3em]{\rule[0.5ex]{#1}{0.55pt}}
\newcommand{\tabOOM}{\multicolumn{3}{c}{\xdash[4em] \textsc{Oom} \xdash[4em]}}

\begin{table*}[h]
	\centering
	\caption{
		Results on the massive graphs listed in Table~\ref{tab:instances}. 
		Here, we compare $\framework_{\kappa=20}$, to \algDefaultFennel{} and \algHashing{}, displaying the fraction of edges cut over total edges (\capCut), running time (\capTime) and memory consumption (\capMemory). 
		Total memory consumption for generated graphs includes memory overheads of the generation process. 
		\algDefaultFennel{} exceeds the available memory of the Raspberry Pi on the trillion-edge graphs.
	}
	\resizebox{0.8\textwidth}{!}{%
		\begin{tabular}{lrrrrrrrrrr}
			& & \multicolumn{3}{c}{$\framework_{\beta=n,\kappa=20}$} & \multicolumn{3}{c}{\algDefaultFennel} & \multicolumn{3}{c}{\algHashing} \\
			\cmidrule(rl){3-5} \cmidrule(rl){6-8} \cmidrule(rl){9-11} 
			$G$ & $k$ & \headCut & \headTime & \headMemory & \headCut & \headTime & \headMemory & \headCut & \headTime & \headMemory \\
			\midrule 
			\multirow{2}{*}{\rotatebox{90}{\instance{gsh15}}} & 4 & 0.078 & 1\,790 & 1.11 & 0.134 & 794 & 4.78 & 0.627 & 142 & 0.004 \\
			& 256 & 0.114 & 5\,428 & 1.11 & 0.200 & 4\,701 & 4.78 & 0.993 & 149 & 0.004 \\
			\midrule
			\multirow{2}{*}{\rotatebox{90}{\instance{cwb12}}} & 4 & 0.041 & 2\,214 & 1.76 & 0.214 & 940 & 5.40 & 0.723 & 143 & 0.004 \\
			& 256 & 0.206 & 5\,661 & 1.76 & 0.381 & 5\,045 & 5.40 & 0.994 & 141 & 0.004 \\
			\midrule
			\multirow{2}{*}{\rotatebox{90}{\instance{uk14}}} & 4 & 0.027 & 2\,081 & 0.26 & 0.182 & 812 & 3.19 & 0.734 & 113 & 0.004 \\
			& 256 & 0.045 & 3\,575 & 0.26 & 0.319 & 3\,958 & 3.19 & 0.995 & 118 & 0.004 \\
			\midrule
			\multirow{2}{*}{\rotatebox{90}{\instance{BA}}} & 4 & $0.337$ & 61\,046 & 4.01 & \tabOOM & 0.374 & 2552 & 4.01 \\
			& 256 & $0.497$ & 143\,634 & 4.03 & \tabOOM & 0.498 & 2561 & 4.01 \\
			\midrule
			\multirow{2}{*}{\rotatebox{90}{\instance{RGG}}} & 4 & $< 0.0001$ & 19\,607 & 2.02 & \tabOOM & 0.377 & 2534 & 2.01 \\
			& 256 & $< 0.0001$ & 72\,354 & 2.03 & \tabOOM & 0.495 & 2532 & 2.01 \\
			\bottomrule
			\vspace{1pt}
		\end{tabular}
	}
	\label{table:huge}
\end{table*}

In this section, we provide experiments on a set of massive graphs shown in Table~\ref{tab:instances}, comparing the performance of $\kappa$\textit{-modified} \linebreak
\hbox{$\framework_{\beta=n,\kappa=20}$}, \textsc{Fennel-Array} and \textsc{Hashing}, run on a Raspberry Pi. Presently, \textsc{Hashing} is the only partitioning method that can partition several billion-node and sparse trillion edge graphs on edge devices. Table~\ref{table:huge} gives detailed per instance results for instances of the massive set.

Our experiments on the massive graphs concur with the results from the Test Set experiments, demonstrating the capabilities of \framework\ under extreme memory constraints. $\framework_{\beta=n,\kappa=20}$ uses less memory and produces higher solution quality than \textsc{Fennel-Array} across all real-world instances with a comparable runtime. Among real-world networks, the largest improvement in memory overhead and solution quality that we observed for \linebreak $\framework_{\beta=n,\kappa=20}$ over \textsc{Fennel-Array} is on \texttt{uk-2014} for $k=256$, for which $\framework_{\beta=n,\kappa=20}$ uses $12.3\times$ less memory than \textsc{Fennel-Array} and improves solution quality by 8.5\%. On generated graphs with up to 17 billion nodes and 1 trillion edges, we demonstrate that \framework\ can partition several billion-node and sparse trillion edge graphs on a Raspberry Pi, with up to 100\% improvement in solution quality over \textsc{Hashing}, while \textsc{Fennel-Array} fails to \hbox{execute under these memory constraints.}

\section{Conclusion}
\label{sec:conclusion}
This paper introduces \framework, a novel framework designed to significantly reduce the memory overhead of streaming graph partitioners through run-length compression of block assignments. \framework\ is highly memory-efficient and can partition massive graphs on edge devices. We provide a modification to the \textsc{la\_vector} bit vector to add append support, thus enabling on-the-fly run-length compression. Compared to \textsc{Fennel} with a standard array, we demonstrate on Test Set instances that \framework\ uses, on average, up to $9.7\times$ less memory with comparable runtime. Our experiments on a set of massive graphs validate the effectiveness of \framework, which can successfully partition graphs with a trillion edges on a Raspberry Pi, demonstrating up to 95.9\% improvement in solution quality over \textsc{Hashing}, the only other method capable of handling such large graphs on edge devices. These results highlight the potential of \framework\ to facilitate high-quality partitioning of large-scale graphs on low-cost, resource-constrained machines, significantly advancing the field of graph processing. Further, our framework is generalizable to other applications of streaming algorithms, which can \hbox{benefit from online run-length compression.} 

\begin{acks}
 We acknowledge support by DFG grant SCHU 2567/5-1.
\end{acks}

\bibliographystyle{ACM-Reference-Format}
\bibliography{compactfixed}

\end{document}